\documentclass[sigconf]{acmart}

\usepackage{booktabs}
\usepackage{enumitem}

\copyrightyear{2026}
\acmYear{2026}
\setcopyright{cc}
\setcctype{by}
\acmConference[FORMATS '26]{1st International Workshop on Data FORMATS for Modern Architectures and Workloads}{May 31-June 05, 2026}{Bengaluru, India}
\acmBooktitle{1st International Workshop on Data FORMATS for Modern Architectures and Workloads (FORMATS '26), May 31-June 05, 2026, Bengaluru, India}
\acmDOI{10.1145/3802514.3812601}
\acmISBN{979-8-4007-2635-4/2026/05}
\begin{document}

\title[Zero-Scan Data Quality]{Zero-Scan Data Quality: Leveraging Table Format Metadata for Continuous Observability at Scale}


\author{Mohit Verma}
\affiliation{\institution{LinkedIn} \city{Bengaluru} \country{India}}
\email{moverma@linkedin.com}

\author{Shantanu Rawat}
\affiliation{\institution{LinkedIn} \city{Bengaluru} \country{India}}
\email{srawat@linkedin.com}

\author{Christian Bush}
\affiliation{\institution{LinkedIn} \city{Sunnyvale} \country{USA}}
\email{chbush@linkedin.com}

\author{Sumedh Sakdeo}
\affiliation{\institution{LinkedIn} \city{Sunnyvale} \country{USA}}
\email{ssakdeo@linkedin.com}

\author{Lokesh Amarnath Ravindranathan}
\affiliation{\institution{LinkedIn} \city{Bengaluru} \country{India}}
\email{lravindranathan@linkedin.com}

\author{Dwarak Bakshi}
\affiliation{\institution{LinkedIn} \city{Bengaluru} \country{India}}
\email{dws@linkedin.com}

\begin{abstract}
Modern table formats such as Apache Iceberg compute and store metadata---commit timestamps, record counts, and column-level statistics such as null counts and value bounds---at write time as part of file writing. These statistics serve query planning, yet they overlap substantially with data quality (DQ) monitoring needs. We describe a metadata-first approach that repurposes write-time statistics for continuous DQ observability---anomaly detection, drift monitoring, null-rate tracking---without scanning any data. Deployed at LinkedIn across 200,000+ Iceberg tables (800+ PB), this approach satisfies approximately 60\% of user-defined DQ rules at zero marginal compute cost and reduces profiling resource consumption by around 50\%. Extending manifest statistics with lightweight counters (sum, zero-value counts, boolean counts) and incrementally mergeable sketches---Theta sketches for distinct counts, KLL sketches for quantiles---can further raise metadata-satisfiable coverage to close to 90\% of production DQ rules. We validate sketch accuracy, mergeability, and storage overhead on production data and propose that table formats should store per-file sketches in Puffin sidecar files, following the same store-then-aggregate pattern used for existing manifest statistics.
\end{abstract}



\maketitle
\renewcommand{\shortauthors}{Verma et al.}

\section{Introduction}

Data quality monitoring is essential for data-intensive organizations, yet the dominant paradigm treats it as a read-time concern: tools such as dbt tests~\cite{dbt}, Great Expectations~\cite{ge}, and Soda Core~\cite{soda} execute SQL queries or full table scans to validate expectations after data is written. At lakehouse scale---hundreds of thousands of tables, petabytes of data, continuous ingestion---ensuring every consumer has the right checks is impractical. To compensate, centralized profiling jobs run asynchronously, but this is prohibitively expensive to operate at scale.

The cost shows up in two ways. First, \emph{scan-based validation}: every existing DQ tool requires reading actual data rows, consuming cluster resources proportional to data volume. In our environment, scan-based profiling for just $\sim$3,000 tables consumed $\sim$1,600~TBhr of compute and read $\sim$1,700~TB from HDFS daily. Second, \emph{detection latency}: because scans run asynchronously, bad data propagates downstream before issues surface.

Modern table formats already solve a significant portion of this problem. Apache Iceberg persists per-data-file column statistics in manifest files---record counts, null counts, NaN counts, bounds, and column sizes~\cite{iceberg}---as byproducts of Parquet/ORC encoding with negligible overhead. Delta Lake stores analogous statistics in its transaction log~\cite{delta}. Yet no existing tool exploits these statistics for quality monitoring. The metadata-first approach benefits organizations of any size: write-time statistics are free regardless of fleet size, and smaller organizations that lack dedicated profiling infrastructure benefit disproportionately.

Our contributions are: (1)~we quantify the overlap between table format metadata and DQ rules---approximately 60\% from manifest statistics alone, rising to $\sim$90\% with counter and sketch extensions (\S\ref{sec:coverage}); (2)~we describe a production architecture that transforms Iceberg commit-time metadata into time-series observability signals; (3)~we report deployment experience at 200,000+ table scale; and (4)~we validate sketch-based extensions on production data and propose concrete format changes.

\section{Motivation}

\subsection{Available Metadata in Table Formats}

Each data file registered in an Iceberg \emph{manifest}---the metadata layer that tracks which files comprise a table snapshot---carries per-file statistics:
\texttt{record\_count}, \texttt{null\_value\_counts},
\texttt{nan\_value\_counts}, \texttt{lower\_bounds},
\texttt{upper\_bounds}, \texttt{value\_counts}, and
\texttt{column\_sizes}~\cite{iceberg}. These are populated
during Parquet or ORC file writing and serialized in Avro manifest
files. Engines expose them via metadata tables---Spark
(\texttt{table.files}, \texttt{table.snapshots}) and Trino
(\texttt{\$files}, \texttt{\$partitions})---primarily for predicate
pushdown, data skipping, and time travel. The same statistics that enable query optimization also enable data quality monitoring---a dual benefit that justifies compute investment at write time~(\S\ref{sec:dual}).
Delta Lake and Apache Hudi store analogous per-file statistics in their transaction log and metadata table, respectively~\cite{delta,hudi}---the pattern is format-general.

\subsection{The Gap: Existing DQ Tools Ignore Format Metadata}

Popular production-grade DQ tools operate by scanning data.
\textbf{dbt tests}~\cite{dbt} use SQL SELECT queries returning failing rows; \textbf{Great Expectations}~\cite{ge} evaluates via SQL or Pandas; \textbf{Databricks}~\cite{databricks} processes complete tables per refresh; and \textbf{Starburst Galaxy}~\cite{starburst} relies on custom SQL rules rather than metadata. Even \textbf{Hudi pre-commit validators}~\cite{hudi}---the closest to write-time DQ---execute SQL queries against staged data before commit, remaining scan-based.

Write-time statistics are computed universally, persisted durably, and queryable cheaply---yet the entire DQ ecosystem ignores them.

\subsection{What Metadata Can and Cannot Do}
\label{sec:coverage}

Manifest statistics are per-file, not pre-aggregated across partitions or tables---combining them requires reading manifest entries. The properties they capture are counts, nulls, and bounds, but not distributional properties (percentiles, distinct counts), referential integrity, or arbitrary SQL predicates.

To quantify coverage, we analyzed $\sim$15,000 user-defined DQ rules authored by data producers across $\sim$1,800 datasets. Separately, data consumers evaluate $\sim$22,000 rules daily across $\sim$8,500 dataset evaluations within their own pipelines. Both populations show a consistent pattern (Table~\ref{tab:coverage}).

\begin{table}[h]
\caption{DQ rule coverage by metadata source.}
\label{tab:coverage}
\footnotesize
\begin{tabular}{lrrll}
\toprule
\textbf{Rule Category} & \textbf{Cons.} & \textbf{Prod.} & \textbf{Example} & \textbf{Source} \\
\midrule
Row-count bounds        & 15\% & 15\% & \texttt{count>1000}      & Manifest \\
Null checks             & 28\% & 26\% & \texttt{notNull(c,<5\%)} & Manifest \\
Min/max, range          & 8\%  & 11\% & \texttt{min(age)>=0}     & Manifest \\
\midrule
\emph{Base manifest}    & \emph{51\%} & \emph{52\%} & & \\
\midrule
Compare metadata\textsuperscript{\dag} & 11\% & 14\% & \texttt{compare(cnt,..)} & Manifest \\
\midrule
\emph{All manifest}     & \emph{62\%} & \emph{67\%} & & \\
\midrule
Sum, mean, zero/true    & 17\% & 8\%  & \texttt{sum(col)>0}      & Counter ext. \\
Distinct count          & 8\%  & 15\% & \texttt{distinct(id)}    & Theta sketch \\
Median, pctl, IQR       & 1\% & $<$1\% & \texttt{median(c)<500}  & KLL sketch \\
\midrule
\emph{Total zero-scan}  & \emph{$\sim$88\%} & \emph{$\sim$90\%} & & \\
\midrule
Expr, regex, enum, arr. & $\sim$12\% & $\sim$10\% & \texttt{expr(a<b*1.15)} & Scan req. \\
\bottomrule
\multicolumn{5}{l}{\scriptsize Cons.\ = consumer rules (downstream pipelines). Prod.\ = producer rules (data owners).} \\
\multicolumn{5}{l}{\scriptsize \textsuperscript{\dag}Compare manifest stats (counts, bounds, nulls) across snapshots/partitions.}
\end{tabular}
\end{table}

Our system already evaluates cross-snapshot comparisons (e.g., \texttt{compare\_count}) from time-series manifest statistics, bringing manifest coverage to $\sim$62--67\%. The ``counter extensions'' row covers metrics such as sum, zero-value count, and boolean counts---not yet in Iceberg manifests but representable as O(1) per-row counters with the same write-time overhead as existing null counts~(\S\ref{sec:counters}). Distinct-count checks---the largest gain for producer rules---are addressable via Theta sketches~(\S\ref{sec:sketches}).

Additionally, $\sim$3,000 user-defined checks monitor data freshness---also surfaceable entirely from Iceberg commit-time metadata. Analyzing historically triggered DQ failures reinforces this: metadata-satisfiable rules account for over 50\% of all consumer-side assertion failures observed in production, driven primarily by row-count and null assertions.

\textbf{Limitations beyond rule coverage.} Four structural gaps remain: (1)~merge-on-read tables (\S\ref{sec:mor}); (2)~wide tables with thousands of columns (\S\ref{sec:columns}); (3)~metrics such as distinct counts and quantiles require sketches (\S\ref{sec:sketches}); (4)~unmaterialized views have no underlying data files---an inherent boundary.

\section{System Architecture}

\begin{figure}[h]
\centering
\includegraphics[width=0.6\columnwidth]{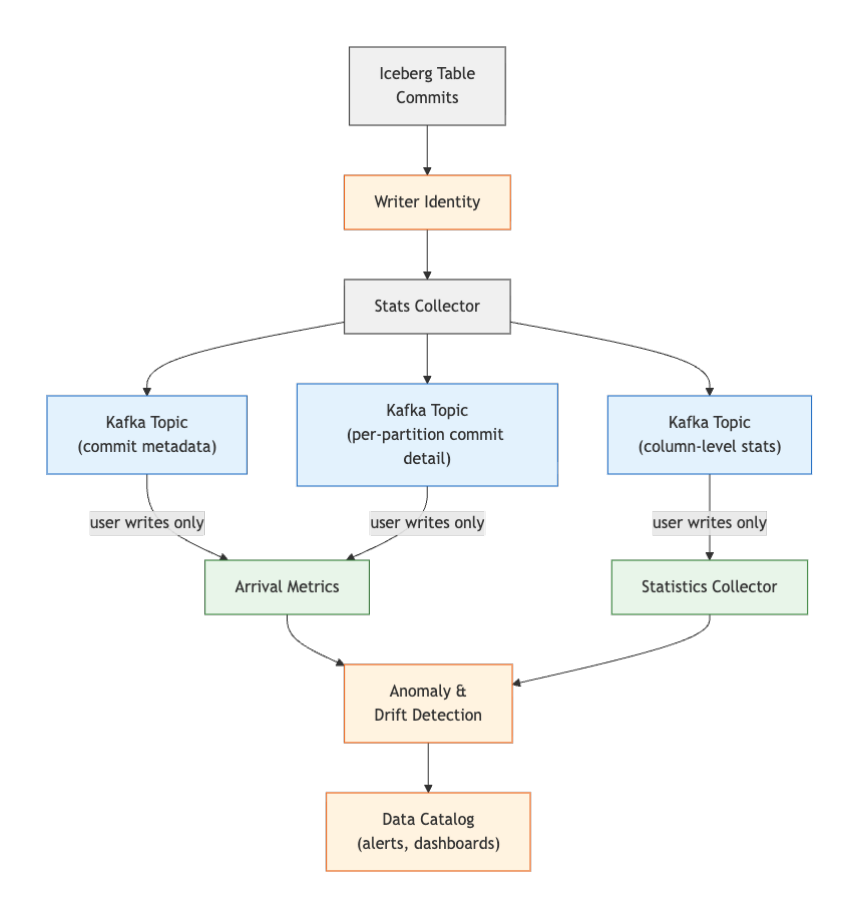}
\caption{Iceberg commit metadata flows through Kafka to consumers that compute arrival and column-level statistics, feeding anomaly and drift detection.}
\label{fig:arch}
\end{figure}

\textbf{Writer identity.} System operations such as compaction, sorting, and purging create new snapshots without changing underlying data. Without distinguishing these from user writes, they add noise to freshness and anomaly signals. We extended Iceberg's commit metadata to include a writer application identity, letting us filter system operations from quality signals.

\textbf{Event collection.} A pre-existing per-table Spark job used for system maintenance was extended to query Iceberg metadata tables---snapshots, manifest entries, data file records---without scanning data rows, and publish event streams to Kafka. Commit timestamps reside in snapshot metadata; column statistics (null counts, bounds) reside in manifest file entries.

\textbf{Consumption and detection.} Two consumers process the streams independently: an arrival pipeline computing per-partition summaries, and a statistics pipeline storing column-level time series. For each table-partition-column combination, we flag row-count anomalies, null-rate drift, range violations, and arrival deviations via the data catalog.\footnote{Our Iceberg extensions and platform code are open source: \url{https://github.com/linkedin/iceberg}, \url{https://github.com/linkedin/openhouse}.}

\section{Production Experience}

\subsection{Resource Impact}

The metadata-first approach is deployed across 200,000+ tables (800+ PB) at LinkedIn. Prior to this, scan-based profiling was limited to $\sim$3,000 opt-in tables due to resource constraints. Replacing scan-based profiling with metrics present in Iceberg metadata reduced both compute and HDFS reads by $\sim$50\% (Table~\ref{tab:resources}). The metadata extraction pipeline itself---reading manifest entries and snapshot metadata across the full fleet---consumes $\sim$13~TBhr of compute per day, < 1\% of the former scan-based profiling cost for 3,000 tables. Tables requiring metrics beyond manifest statistics (percentiles, distinct counts) and unmaterialized views still run full scans; the extensions proposed in \S\ref{sec:future} aim to close this gap.

\begin{table}[h]
\caption{Resource consumption before and after metadata-first profiling.}
\label{tab:resources}
\small
\begin{tabular}{lrrr}
\toprule
\textbf{Metric} & \textbf{Before} & \textbf{After} & \textbf{Change} \\
\midrule
Compute (TBhr/day)  & 1,623 & 796   & $-$51\% \\
HDFS reads (TB/day) & 1,669 & 913   & $-$45\% \\
\bottomrule
\end{tabular}
\end{table}
\subsection{Detection Latency and Coverage}

Manifest statistics are available at commit time, so quality signals are generated within minutes (p50 $<$5~min, p99 $<$20~min) versus 2--24 hours for scheduled scan-based profiling. Coverage expanded from $\sim$3,000 opt-in tables to the full 200,000+ table fleet at near-zero marginal cost---quality monitoring for 197,000+ previously unobserved tables.

\subsection{Incidents and Operational Insights}

\textbf{Freshness gap with real-world impact.} A legacy ETL flow silently began writing invalid data to a business-critical dimension table, setting all records to inactive and dropping $\sim$3\% of rows. The corruption impacted 17 downstream applications and external customer reports. The incident post-mortem noted: ``no data freshness alerts configured.'' Our metadata-first arrival monitoring---which tracks commit timestamps across the full table fleet at near-zero cost---would have surfaced the anomaly within minutes of the first corrupted write.

\textbf{Null-count spike in feature store.} A feature engineering pipeline exhibited a sharp increase in null percentage over three days in a job embedding feature store, degrading downstream ML model quality. Null-count monitoring via null counts stored in Iceberg manifest files would have surfaced the anomaly on the first day, rather than the manual discovery three days later that triggered the incident.

\textbf{Fleet-wide operational insights.} Metadata-first monitoring at near-zero marginal cost has already revealed patterns invisible at data lake scale:

\begin{itemize}[leftmargin=*,nosep]
\item \emph{Lookback writes}: critical datasets frequently rewrite recent partitions, creating incomplete intermediate states visible to concurrent consumers. We surface this as a data-readiness signal.
\item \emph{Over-partitioning}: tables with thousands of small partitions per day---an anti-pattern causing metadata bloat and query planning overhead. Fleet-level visibility led to targeted compaction recommendations.
\item \emph{Maintenance noise}: compaction and sorting operations register as data changes. The writer identity extension described in \S3 was essential for filtering these.
\item \emph{Arrival variability}: even nominally ``daily'' tables exhibit high variance in commit timing. Without continuous metadata monitoring, consumers cannot distinguish normal delays from failures.
\end{itemize}

\section{Proposed Extensions and Validation}
\label{sec:future}

Our experience suggests that table formats should treat data quality as a first-class design concern. Below we propose extensions and present experimental validation on production data.

\subsection{Lightweight Counter Extensions}
\label{sec:counters}

Rules checking column sums, zero-value counts, boolean true/false counts, empty-string counts, and mean values account for 8--17\% of production DQ rules (Table~\ref{tab:coverage}). These are all representable as O(1) per-row aggregates accumulated during the same encoding pass that already produces null counts. Null counts are ``free'' today because the file format computes them as a byproduct of definition-level encoding and stores them in the column chunk footer; writers copy them into Iceberg manifest entries at commit time. 
The proposed counters---\texttt{sum}, \texttt{zero\_value\_counts}, and \texttt{true\_value\_counts}---require a four-layer integration: (1)~defining fields in Parquet/ORC footers; (2)~updating engines (Spark, Trino) to populate them during encoding; (3)~extending Iceberg manifest schemas; and (4)~modifying manifest-writing code to propagate these values. Each counter adds a single 8-byte field per column per data file, imposing negligible overhead. The primary barrier is ecosystem coordination rather than compute or storage cost. Mean is derived from sum and record count, requiring no additional field.

\subsection{Incrementally Mergeable Sketches}
\label{sec:sketches}

Beyond simple counters, distinct-count and distribution checks require probabilistic data structures. Two sketch types address this:

\begin{itemize}[leftmargin=*,nosep]
  \item \textbf{Theta sketch} (NDV): approximate distinct count. 33~KB per high-cardinality column, O(1) update per row, $<$3\% relative error~\cite{datasketches}.
  \item \textbf{KLL sketch} (quantiles): approximate percentiles. 5~KB per column, $<$1.65\% rank error with $k$=200.
\end{itemize}

The key property is \emph{incremental mergeability}: file-level sketches can be combined across files with bounded error, preserving the zero-scan property. For Theta sketches, the union operation is mathematically lossless---merging per-file sketches produces an NDV estimate identical to a single-pass computation over all data.

\textbf{Experimental validation.} We built per-file Theta and KLL sketches for a production table ($\sim$100\,M rows, 20 columns, 100 data files), merged them into table-level sketches, and compared
the results against exact values obtained by full scan. Table~\ref{tab:sketch} summarizes the results.

\begin{table}[h]
\caption{Sketch validation on production data ($\sim$100\,M rows, 100 files).}
\label{tab:sketch}
\small
\begin{tabular}{llr}
\toprule
\textbf{Sketch} & \textbf{Metric} & \textbf{Result} \\
\midrule
Theta (NDV) & Relative error     & $<$0.5\% \\
Theta (NDV) & Merge 100 per-file sketches & lossless (0\%) \\
Theta (NDV) & Storage per column per data file  & $\leq$33~KB \\
Theta (NDV) & Merge cost (100 files $\to$ 1) & 3--7~seconds \\
Theta (NDV) & Total storage (20 cols, 100 files) & 20.7~MB \\
\midrule
KLL (quantiles) & Relative error & $<$1\% \\
KLL (quantiles) & Merge 100 per-file sketches & $<$1\% error \\
KLL (quantiles) & Storage per column per data file  & $\sim$5~KB \\
\bottomrule
\end{tabular}
\end{table}

The observed Theta error ($<$0.5\%) is well within the theoretical $<$3\% bound~\cite{datasketches}; Theta union is mathematically lossless---merging per-file sketches yields an estimate identical to a single-pass computation. For KLL, merged quantiles remain
within the bounded-error guarantee ($\epsilon \approx 1.65\%$ rank error for $k$=200).

\textbf{Storage and merge cost.} Theta sketches cap at $\sim$33~KB per column per data file regardless of cardinality (4,096 entries $\times$ 8~bytes); low-cardinality columns use 32--120~bytes. Total storage grows with columns $\times$ files, so sketch computation should be limited to high-value columns~(\S\ref{sec:columns}). Merging 100 per-file Puffin files into one table-level file took 3--7~seconds. Adding this to the write path is undesirable, which motivates storing sketches per data file and deferring merge to read time.

A natural design question is where sketches should reside. Embedding them in Parquet footers would make them automatic but it inflates the footers read by every query planner, not just DQ consumers. We recommend \textbf{per-data-file Puffin sidecar files}~\cite{puffin}: opt-in, zero overhead on the query-planning path, and compatible with any underlying file format. Aggregation follows the same store-then-aggregate pattern Iceberg already uses for manifest statistics---per-file sketches are merged (Theta union, KLL merge) at read time.

\subsection{Merge-on-Read and Delete Vectors}
\label{sec:mor}

In merge-on-read (MOR) mode, manifest statistics for base data files
  do not reflect pending deletes. MOR tables arise almost exclusively
  from change-data-capture (CDC) pipelines that replicate upstream
  database mutations into the lakehouse. In our deployment (Iceberg
  format~V2), < 1\% of tables use MOR with positional deletes;
  the remaining 99\% use copy-on-write, where statistics are exact at
  write time. Even for this narrow MOR fraction, metadata-based DQ
  remains viable through two mechanisms.

  \textbf{(1)~Compaction restores exact statistics.} Scheduled
  compaction merges base data files with delete files, producing new
  data files with correct manifest statistics. Post-compaction, a MOR
  table is indistinguishable from a COW table for metadata-based DQ. In
  practice, CDC pipelines trigger compaction after every
  \texttt{merge-into} statement, so the window of stale metadata is
  typically seconds to minutes.

  \textbf{(2)~Bounded statistics bridge the compaction gap.} Even
  within that window, delete manifests track each delete file's record
  count. For \emph{position deletes} (and deletion vectors in
  format~V3), the correction is exact:
  $N_{\text{actual}} = N - N_{\text{pos}}$.

  Concretely, a production partition with 1\,M rows and 50\,K position
  deletes yields an exact corrected count of 950{,}000---rules such as
  \texttt{count~>~500K} resolve from metadata alone, with no scan
  required.

\subsection{Declarative Quality Constraints}

  Relational databases enforce column-level \texttt{CHECK} constraints
  at write time; lakehouse table formats have no equivalent~\cite{icebergcheck}.  We
  propose \emph{aggregate quality constraints}: expressions over
  manifest statistics that Iceberg evaluates during the commit protocol,
  before a snapshot becomes visible to readers.

  A constraint is stored as table-level metadata (analogous to sort
  orders or partition specs) and takes the form of a predicate over
  per-file or per-partition aggregates already present in the manifest.
  Examples:

  \noindent
  \texttt{null\_count(user\_id) / record\_count < 0.05}\\
  \texttt{record\_count > 1000}\\
  \texttt{max(event\_ts) >= commit\_ts - interval '2 hours'}

  \noindent Because these predicates reference only manifest-level
  statistics, enforcement adds zero scan cost and applies uniformly
  across every write engine---Spark, Trino, Flink, or any ETL
  framework---that commits through the Iceberg API. A commit that
  violates a constraint is rejected with a descriptive error before the
  snapshot is published, preventing bad data from reaching downstream
  consumers. 

\subsection{Column Prioritization for Wide Tables}
\label{sec:columns}

Wide tables with thousands of columns make collecting per-column statistics expensive. In our deployment, Iceberg manifest entries are capped at 200 columns per data file; sketch computation faces the same scaling concern. At 33~KB per high-NDV column, 500 columns $\times$ 100 files $=$ 1.65~GB of per-file sketch metadata ($\sim$2\% of a typical 80~GB table). Column-level lineage metadata---tracking which columns downstream consumers actually read---will help to prioritize statistics for high-impact columns, keeping both manifest and sketch overhead bounded.

\subsection{Dual Benefit: Quality and Query Optimization}
\label{sec:dual}

Several proposed statistics serve both DQ monitoring and query optimization, strengthening the case for write-time computation:

\footnotesize
\begin{tabular}{lll}
\toprule
\textbf{Statistic} & \textbf{DQ Use} & \textbf{Query Opt.\ Use} \\
\midrule
Theta (NDV)   & Cardinality drift    & Join-order estimation \\
KLL (quantiles) & Distribution checks & Histogram cost est. \\
Counters (sum) & Aggregate bounds     & Selectivity est. \\
\bottomrule
\end{tabular}

\normalsize
Statistics computed once at write time benefit both read-time query planning and continuous DQ monitoring.

\section{Related Work}

Data profiling has a long history~\cite{profiling}. SPADE~\cite{spade} synthesizes DQ assertions for LLM pipelines but remains query-based. Commercial formats like Delta Lake~\cite{delta} and Hudi~\cite{hudi} support row-level \texttt{CHECK} constraints or pre-commit validators, yet both enforce quality through active data access. General lakehouse architectures~\cite{lakehouse} do not address metadata-driven observability. While Apache DataSketches~\cite{datasketches} provides the Theta~\cite{kmv} and KLL~\cite{kll} algorithms we validated, and Iceberg's Puffin~\cite{puffin} supports sketch storage, existing implementations require retroactive full-table scans rather than inline write-time computation. To our knowledge, our approach is the first to use manifest-level statistics as the sole input for DQ monitoring at production scale.

\bibliographystyle{ACM-Reference-Format}
{

}

\end{document}